\documentstyle[preprint,aps,pra,epsfig,cite]{revtex}
\begin{document}

\draft
\title{Entangled state based on nonorthogonal state
}
\author{Osamu Hirota, and Masaaki Sasaki$^\dagger$\\
Research Center for Quantum Communications, \\
Tamagawa University, Machida, Tokyo, Japan.\\
$^\dagger$ Communication Research Laboratory, Tokyo, Japan.
}

\maketitle

\begin{abstract}
Properties of entangled states based on nonorthogonal states
are clarified. Especially, it is shown that they can have 
complete degree of entanglement.

\end{abstract}


\section{INTRODUCTION}
Entanglement and its information theoretic aspects have been
studied by many authors
\cite{Bennett,Bennett2,vedral,Wootters,Wootters2}.
For a pure entangled state of a bipartite system
$|\rho_{AB}\rangle$, a measure of entanglement is defined as
\cite{Bennett, Barnett}
\begin{equation}
E(|\rho_{AB}\rangle)=-{\rm Tr}_{A}{\rho}_{A}\log {\rho}_{A},
\quad
{\rho}_{A}={\rm Tr}_{B}|\rho_{AB}\rangle\langle\rho_{AB}|,
\end{equation}
which is called as "entropy of entanglement".
This quantity enjoys two kinds of information theoretic
interpretations.
One of them is entanglement of formation which means
the asymptotic number $k$ of standard singlet required to
locally prepare faithfully $n$ identical copies of
a system in bipartite state $|\rho_{AB}\rangle$ for
very large $k$ and $n$.
Other is distillable entanglement which means the
asymptotic number of singlets $k$
that can be distilled from $n$ identical copies of
$|\rho_{AB}\rangle$.
In particular, it satisfies
\begin{equation}
\lim _{n, k \rightarrow \infty} \frac{k}{n} = E(|\rho_{AB}\rangle).
\end{equation}
Explicit expressions for $E(|\rho_{AB}\rangle)$ is only known
in the case of two qubit systems \cite{Wootters}.
In fact, it is given as
\begin{equation}
E(|\rho_{AB}\rangle)= H[\frac{1}{2}(1+\sqrt {1 - C(\rho_{AB})^2})]
\end{equation}
where $H[x]$ is the entropy function and
$C(\rho_{AB})$ is the "concurrence" defined by
$C(\rho_{AB})=|\langle\rho_{AB}|\tilde\rho_{AB}\rangle|$ with
$|\tilde\rho_{AB}\rangle=\sigma|\rho_{AB}\rangle^*$.
The similar analytic formulas for mixed states of qubits is
also obtained for the properly defined entanglement of formation
\cite{Wootters2}.
In this paper, we study properties of entangled states
based on nonorthogonal states such as coherent states.
An implementation scheme for manipulating such states are also
discussed.

\section{QUASI BELL STATE}
\subsection{General definition}
Let us define the entangled state based on nonorthogonal states
$|\psi_1 \rangle$ and $|\psi_2 \rangle$ such as
$ \langle \psi_1 | \psi_2 \rangle = \kappa$ and
$\langle \psi_2 | \psi_1 \rangle = \kappa^*$.
They can be described by
\begin{eqnarray}
\left\{
\begin{array}{lcl}
|\Psi_1 \rangle &=& h_{1} (|\psi_1 \rangle_A|\psi_2 \rangle_B
+|\psi_2 \rangle_A|\psi_1 \rangle_B ) \\
|\Psi_2 \rangle &=& h_{2} (|\psi_1 \rangle_A|\psi_2 \rangle_B
-|\psi_2 \rangle_A|\psi_1 \rangle_B )\\
|\Psi_3 \rangle &=& h_{3} (|\psi_1 \rangle_A|\psi_1 \rangle_B
+ |\psi_2 \rangle_A|\psi_2 \rangle_B )\\
|\Psi_4 \rangle &=& h_{4} (|\psi_1 \rangle_A|\psi_1 \rangle_B
-|\psi_2 \rangle_A|\psi_2 \rangle_B)
\end{array}
\right.
\end{eqnarray}
where
$\{h_{i}\}$ are normalized constant:$h_{1}=h_{3}=1/\sqrt{2(1+\kappa^{2})}$,
$h_{2}=h_{4}=1/\sqrt{2(1-\kappa^{2})}$.
They are not orthogonal each other.
Here, if $\kappa = \kappa^*$, then the Gram matrix of them
becomes very simple as follows:
\begin{equation}
G=
\left(
\begin{array}{cccc}
1& 0& D& 0\\
0& 1& 0& 0\\
D& 0& 1& 0\\
0& 0& 0& 1\\
\end{array}
\right)
\label{ohmsgrammat}
\end{equation}
where $D=\frac{2 \kappa}{1 + {\kappa}^2}$.
If the basic states are orthogonal, then they are Bell states.
Let us discuss the entropy of entanglement for the above states.
We, first, calculate the reduced density operators of
the quasi Bell state.
They are ${{\rho}_{A}}^{(1)}={{\rho}_{A}}^{(3)}$ and
${{\rho}_{A}}^{(2)} = {{\rho}_{A}}^{(4)} $.
Their concrete forms are
\begin{equation}
{{\rho}_{A}}^{(1)}
 = \frac{1}{2(1+\kappa^2)}
\{|\psi_1 \rangle_A\langle\psi_1| +
\kappa |\psi_1 \rangle_A\langle \psi_2|
+\kappa |\psi_2 \rangle_A\langle \psi_1|
+ |\psi_2 \rangle_A\langle \psi_2\}
\end{equation}
\begin{equation}
{{\rho}_{A}}^{(2)}=\frac{1}{2(1-\kappa^2)}
\{|\psi_1 \rangle_A\langle \psi_1|
-\kappa |\psi_1 \rangle_A\langle \psi_2|
-\kappa |\psi_2 \rangle_A\langle \psi_1|
+ |\psi_2 \rangle_A\langle \psi_2|\}
\end{equation}
The eigenvalues of the above density operators
${{\rho}_{A}}^{(1)}$(or ${{\rho}_{A}}^{(3)}$) are given as follows
by using the Gram matrix elements
$C_{ij}=|\langle\Psi_i|\Psi_j\rangle|$ of Eq. (\ref{ohmsgrammat}):
\begin{equation}
\lambda_{1/1} = \frac{(1+\kappa)^2}{2(1+\kappa^2)}= \frac{1+C_{13}}{2},
\quad
\lambda_{2/1} = \frac{(1-\kappa)^2}{2(1+\kappa^2)} = \frac{1-C_{13}}{2}
\end{equation}
and for ${{\rho}_{A}}^{(2)}$(or ${{\rho}_{A}}^{(4)}$), we have
\begin{equation}
\lambda_{1/2} =\frac{1}{2}= \frac{1+C_{24}}{2},
\quad
\lambda_{2/2} = \frac{1}{2}= \frac{1-C_{24}}{2}.
\end{equation}

The entropy of entanglement is then
\begin{equation}
E(|\Psi_1\rangle)=E(|\Psi_3\rangle)
                 = - \frac{1+C_{13}}{2} \log \frac{1+C_{13}}{2}
                   - \frac{1-C_{13}}{2} \log \frac{1-C_{13}}{2}
\end{equation}
and $E(|\Psi_2\rangle)=E(|\Psi_4\rangle)=1$.
Thus $|\Psi_2\rangle$ and $|\Psi_4\rangle$ have perfect entanglement,
even though the enatangled states consist of nonorthogonal state in
each subsystem. These results are true for arbitrary nonorthogonal states
with $\langle \psi_1 | \psi_2 \rangle$ $=$ $\langle \psi_2 | \psi_1
\rangle$ $=$ $\kappa$ and do not depend on the physical dimension of
the systems.

\subsection{Generation of quasi Bell states}
It is well known that an entangled state can be generated by
Walsh-Hadamard gate and CN gate. That is, when the input state
for control bit is a superposition state generated
by Walsh-Hadamard gate, the output state of the CN gate is
an entangled state.
The W-H gate(Walsh-Hadamard transformation) is described by
\begin{equation}
U_{WH} = \exp\{\theta (|0 \rangle \langle 1 | - |1 \rangle \langle 0 |)\}
\end{equation}
and the unitary operator for the control NOT is
\begin{equation}
U_{CN}=| 0 \rangle_C \langle 0 | \otimes I_T +
| 1 \rangle_C \langle 1 | \otimes
(| 0 \rangle_T\langle 1 | + | 1 \rangle_T\langle 0 |)
\end{equation}
The function of the control NOT(CN gate) is as follows:
\begin{eqnarray}
\left\{
\begin{array}{lcl}
|0 \rangle_C |0 \rangle_T &\rightarrow& U_{CN}|0 \rangle_C |0 \rangle_T
=|0 \rangle_C |0 \rangle_T\\
|0 \rangle_C |1 \rangle_T &\rightarrow& U_{CN}|0 \rangle_C |1 \rangle_T
=|0 \rangle_C |1 \rangle_T\\
|1 \rangle_C |0 \rangle_T &\rightarrow& U_{CN}|1 \rangle_C |0 \rangle_T
=|1 \rangle_C |1 \rangle_T\\
|1 \rangle_C |1 \rangle_T &\rightarrow& U_{CN}|1 \rangle_C |1 \rangle_T
=|1 \rangle_C |0 \rangle_T
\end{array}
\right.
\end{eqnarray}
where $C$ and $T$ mean control mode, and target mode,
respectively.

On the two state space spanned by nonorthogonal states:
$|\psi_1 \rangle$ and $|\psi_2 \rangle$,
we can consider general scheme to manipulate the quasi
Bell states. Let us define the orthonormal basis
\begin{eqnarray}
|\psi_e \rangle &=& (|\psi_1 \rangle + |\psi_2 \rangle)/
\sqrt{2(1+\kappa)},\label{even_odd basis-1}\\
|\psi_o \rangle &=& (|\psi_1 \rangle - | \psi_2 \rangle)/
\sqrt{2(1-\kappa)}.
\label{even_odd basis-2}
\end{eqnarray}
They play a role of qubit basis $\{|0\rangle, |1\rangle\}$.
We can then go along with quantum logic operations on qubit systems. In
terms of the basis defined by
Eq. (\ref{even_odd basis-1}),
and Eq. (\ref{even_odd basis-2}),
the required gates are
\begin{equation}
U_{WH} = \exp\{\theta (|\psi_e \rangle \langle \psi_o | -
|\psi_o \rangle \langle \psi_e |)\}
\end{equation}
\begin{eqnarray}
U_{CN}&=&|\psi_e \rangle_C \langle \psi_e | \otimes I_T
\nonumber\\
&+&
| \psi_o \rangle_C \langle \psi_o | \otimes
\Big(| \psi_e \rangle_T\langle \psi_o |
+ | \psi_o \rangle_T\langle \psi_e |\Big)
\end{eqnarray}
The W-H gate acts on the input superposition state as follows:
\begin{eqnarray}
|\psi_e \rangle_C &\rightarrow& U_{WH} |\psi_e \rangle_C\nonumber\\
&=&|\psi_e \rangle_C+|\psi_o \rangle_C
\end{eqnarray}
Thus we have
\begin{eqnarray}
K\Big(|\psi_e \rangle_C&+&|\psi_o \rangle_C \Big) |\psi_e \rangle_T
\nonumber \\
&\rightarrow&
U_{CN}K\Big(|\psi_e \rangle_C+|\psi_o \rangle_C \Big)
|\psi_e \rangle_T 	\nonumber \\
&=&
K\Big(|\psi_e \rangle_C |\psi_e \rangle_T +
|\psi_o \rangle_C |\psi_o \rangle_T\Big)
\nonumber\\
&=&
h_{3} \Big(|\psi_1 \rangle_C|\psi_1 \rangle_T +
|\psi_2 \rangle_C|\psi_2 \rangle_T \Big)
\end{eqnarray}
where $K$ is the normalized constant. The final state is
one of quasi Bell state.
Thus we have quasi Bell states
based on such operations.
If we use the coherent states as the basic states,
then the above gates correspond to bosonic gates
whose realization is discussed in the later section.

\subsection{General case}
Here let us consider the general pure entangled state of nonorthogonal state.
They can be described as follows:
\begin{eqnarray}
\left\{
\begin{array}{lcl}
|\Psi_1 \rangle &=& g_{1} (\beta |\psi_1 \rangle_A|\psi_2 \rangle_B
+\sqrt{1-\beta^2}|\psi_2 \rangle_A|\psi_1 \rangle_B ) \\
|\Psi_2 \rangle &=& g_{2} (\beta|\psi_1 \rangle_A|\psi_2 \rangle_B
-\sqrt{1-\beta^2}|\psi_2 \rangle_A|\psi_1 \rangle_B )\\
|\Psi_3 \rangle &=& g_{3} (\beta|\psi_1 \rangle_A|\psi_1 \rangle_B
+ \sqrt{1-\beta^2}|\psi_2 \rangle_A|\psi_2 \rangle_B )\\
|\Psi_4 \rangle &=& g_{4} (\beta|\psi_1 \rangle_A|\psi_1 \rangle_B
-\sqrt{1-\beta^2}|\psi_2 \rangle_A|\psi_2 \rangle_B )
\end{array}
\right.
\end{eqnarray}
where $g_{i}$ is normalized constant, and $\beta$ is real number.
Since all the elements of the Gram matrix is not zero, they are
not orthogonal states.
The reduced density operators in this case become
${{\rho}_{A}}^{(1)}={{\rho}_{A}}^{(3)}$ and
${{\rho}_{A}}^{(2)} = {{\rho}_{A}}^{(4)} $, and
\begin{eqnarray}
\rho_{A}^{(1)}&=& k_1
\{\beta^2|\psi_1 \rangle_A\langle\psi_1| +
\kappa\beta\sqrt{1-\beta^2}|\psi_1 \rangle_A\langle \psi_2 \nonumber\\
&+&\kappa\beta\sqrt{1-\beta^2}|\psi_2 \rangle_A\langle \psi_1|
+(1-\beta^2)|\psi_2 \rangle_A\langle \psi_2 |\}\\
\rho_{A}^{(2)}&=&k_2
\{\beta^2|\psi_1 \rangle_A\langle \psi_1 |
-\kappa\beta\sqrt{1-\beta^2}|\psi_1 \rangle_A\langle \psi_2 |\nonumber\\
&-&\kappa\beta\sqrt{1-\beta^2}|\psi_2 \rangle_A\langle \psi_1|
+(1-\beta^2)|\psi_2 \rangle_A\langle \psi_2 |\}
\end{eqnarray}
where $k_i= \frac{1}{1 \pm 2{\kappa}^2\beta\sqrt{1-\beta^2}}$ is
normalized constant. The analysis in this case is also easy.
Here we again assumed that $\langle \psi_1 | \psi_2 \rangle$
$=$ $\langle \psi_2 | \psi_1 \rangle$ $=$ $\kappa$.

\section{QUASI BELL STATES OF COHERENT STATES}
Let us consider the binary coherent states of a bosonic mode
$\{|\alpha\rangle, |-\alpha\rangle\}$, where
$(\kappa=\langle\alpha|-\alpha\rangle={\rm e}^{-2|\alpha|^2})$.
Then the quasi Bell states are
\begin{eqnarray}
\left\{
\begin{array}{lcl}
|\Psi_1 \rangle &=& h_{1} (|\alpha \rangle_A|-\alpha \rangle_B
+|-\alpha \rangle_A|\alpha \rangle_B ) \\
|\Psi_2 \rangle &=& h_{2} (|\alpha \rangle_A|-\alpha \rangle_B
-|-\alpha \rangle_A|\alpha \rangle_B )\\
|\Psi_3 \rangle &=& h_{3} (|\alpha \rangle_A|\alpha \rangle_B
+ |-\alpha \rangle_A|-\alpha \rangle_B )\\
|\Psi_4 \rangle &=& h_{4} (|\alpha \rangle_A|\alpha \rangle_B
-|-\alpha \rangle_A|-\alpha \rangle_B)
\end{array}
\right.
\end{eqnarray}
where $\alpha$ is coherent amplitude of light field.
The average photon numbers of the reduced states are
\begin{equation}
\langle n_{A}^{(1)} \rangle = \frac{(1-\kappa^2)}{(1+\kappa^2)} |\alpha |^2,
\quad
\langle n_{A}^{(2)} \rangle = \frac{(1+\kappa^2)}{(1-\kappa^2)} |\alpha |^2
\end{equation}
Thus the quasi Bell states can have arbitrary photon,
and approach to the Bell states as $|\alpha|\rightarrow\infty$.
We mention the characteristic function of quasi Bell states
defined as
\begin{eqnarray}
C(\xi, \eta) &=& {\rm Tr}\big[
   |\Psi\rangle\langle\Psi|
       \exp({\xi} {\mit a}_A^\dagger)
       \exp({-\xi^*} {\mit a}_A)
       \exp({\eta} {\mit a}_B^\dagger)
       \exp({-\eta^*} {\mit a}_B)
                           \big] \nonumber \\
&\times& \exp\{-(|\xi |^2 + |\eta |^2)/2\}
\end{eqnarray}
where $a$ and $a^\dagger$ are the annihilation and
creation operators, respectively.
They are actually
\begin{eqnarray}
C(\xi, \eta |i=1, 2)&=&{h_{i}}^2\exp\{-(|\xi |^2 +
|\eta |^2)/2\}\{\exp(A_1-B_1)\alpha\nonumber\\
&+&\exp(-A_1+B_1)\alpha \pm \exp(A_2 - B_2)\alpha \nonumber \\
&\pm& \exp(-A_2 + B_2)\alpha\}\\
C(\xi, \eta |i=3,4)&=&{h_{i}}^2\exp\{-(|\xi |^2 +
|\eta |^2)/2\}\{\exp(A_1+B_1)\alpha\nonumber \\
&+&\exp(-A_1-B_1)\alpha \pm \exp(A_2 +B_2)\alpha \nonumber \\
&\pm& \exp(-A_2 - B_2)\alpha\}
\end{eqnarray}
where $A_1 = (\xi - \xi^*), A_2 = (\xi + \xi^*), B_1 = (\eta - \eta^*),
B_2 = (\eta + \eta^*)$.
It is worthy to mention that the quasi Bell states do not
belong the Gaussian state in contrast to that two mode squeezed state
does so.

\section{PHYSICAL REALIZATION}
In order to manipulate the quasi Bell states of bosonic coherent states,
one needs quantum gates acting on a state space spanned by the
relevant coherent states.
A convenient basis is the even and odd coherent
states.
Let us denote them as $\{ \vert e\rangle , \vert o\rangle \}$ hereafter.
It would be much difficult to realize quantum gates for these
{\it macroscopic qubits}.
Cochrane, Milburn, and Munro proposed a physical model for such gates
\cite{Cochrane}.
In their model, a CN gate is made by applying an $H$ gate to
the mode $a$, the target bit, then coupling the target and the control
(the mode $b$), and finally applying another $H$ gate to the target again.
This CN gate operation is actually valid in a certain limited
case of the coherent state amplitude $\alpha$ and the classical field
amplitude $\gamma$. However, their model is indeed indicative.
If the $H$ gate is capable of generating the Schr\"odinger cat state of
coherent state, that is, includes an appropriate nonlinear Hamiltonian
instead of the linear interaction of Eq. \cite{Cochrane},
then their scheme provides the universal bosonic gates.
(This was also recognized by Tatsuta et al. \cite{Tatsuta})
Concerning the two bit interaction, the cross Kerr effect always suffices.
Thus the problem is to synthesize the appropriate nonlinear Hamiltonian.

One bit gate operations for macroscopic bosonic qubits can be represented
by rotations on the two-state space. For example, the Hadamard gate is
represented by
\begin{equation}
\hat U_H=-i{\rm exp}[i{\pi\over2}\hat Q]{\rm exp}[{\pi\over4}\hat P],
\label{U_H}
\end{equation}
\begin{equation}
\hat P=\vert e\rangle\langle o\vert - \vert o\rangle\langle e\vert, \quad
\hat Q=\vert e\rangle\langle e\vert - \vert o\rangle\langle o\vert,
\label{P&Q}
\end{equation}
The physical process corresponding to $\hat U_H$ includes essentially
multiphoton nonlinear precess. The corresponding Hamiltonians were
studied by Sasaki and Hirota \cite{Sasaki96}.

The case of our interest is when the amplitude $\vert\alpha\vert$ is
small, in which the assumption used in \cite{Cochrane}.
Then the Hamiltonian
$\hat P$ and $i\hat Q$ can be effected by the nonlinear Hamiltonian
including finite number of nonlinearity.
For mathematical convenience, we consider the Hamiltonian for
$\tilde U_H=\hat D(-\alpha) \hat U_H \hat D(\alpha)$.
First we introduce a cut off photon number $M$ for a weak
coherent state such that its photon number distribution in $n>M$
becomes negligibly small. Second define
\begin{equation}
\hat P_M=-2\sqrt{1-c_0^2}\left(
         \sum_{l=0}^M {\frac {(-{\hat a}^\dagger)^l{\hat a}^l}{l!}}
            \sum_{n=1}^M d_n {\frac{\hat a^{n}} {\sqrt{n!}}}
            - {\rm h. c.}\right),
\label{P:Hamiltonian}
\end{equation}
\begin{equation}
\hat Q_M=4c_0\sum_{l=0}^M {\frac {(-{\hat a}^\dagger)^l{\hat a}^l}{l!}}
        +2\sqrt{1-c_0^2}\left(
         \sum_{l=0}^M {\frac {(-{\hat a}^\dagger)^l{\hat a}^l}{l!}}
            \sum_{n=1}^M d_n {\frac{\hat a^{n}} {\sqrt{n!}}}
            + {\rm h. c.}\right),
\label{Q:Haniltonian}
\end{equation}
where
\begin{equation}
c_n={\rm e}^{-2\alpha^2}{{(-2\alpha)^2}\over{\sqrt{n!}}}, \quad
d_n={{c_n}\over{\sqrt{\sum_{n=1}^M}c_n^2}}.
\end{equation}
Then $\tilde U_H$ can be represented
\begin{equation}
\tilde U_H=-i{\rm exp}[i{\pi\over2}\hat Q_M]{\rm exp}
                      [{\pi\over4}\hat P_M]+O(\delta_M),
\end{equation}
where
\begin{equation}
\delta_M=1-{{\sum_{n=1}^Mc_n^2}\over{1-c_0^2}}.
\end{equation}
The Hamiltonian of $\hat P_M$ and $\hat Q_M$ still seem to be unrealistic.
One possible way to make realistic is to decompose them into a cascade
process of lower order nonlinear processes. In fact, as suggested by
Harel and Akulin \cite{Harel-Akulin}, and
Lloyd and Braunstein\cite{Lloyd-Braunstein}, it is possible in principle
to synthesize the required unitary dynamics by lower order nonlinear
Hamiltonians. In particular, it can be shown that the nonlinearity up to
third order and the cross Kerr nonlinearity suffice to implement
the universal gates for macroscopic bosonic qubits.

\section{CONCLUSION}
We investigated properties of entangled states based on nonorthogonal state
such as coherent states.
Implementation of quantum gates for such macroscopic qubits was suggested.
We would like to find more simple generation method of such a quasi Bell
states.

\begin{acknowledgments}
We are grateful to M.Ban, S.Barnett, C.Bennett,
S. J. van Enk, C.Fuchs, A.Holevo, K.Kato,
R.Jozsa, M.Osaki, and P.Shor for helpful discussions.
\end{acknowledgments}

\end{document}